\documentclass[notitlepage,aps,pra,twocolumn,groupaddress]{revtex4-1}

\usepackage{stmaryrd}
\usepackage{amssymb,amsmath,amsthm,amsfonts,amsbsy}
\usepackage{bm,bibunits,color,chngcntr,epsfig,epstopdf,graphicx,dsfont}
\usepackage{hyperref,lipsum,,makecell,mathrsfs,rotating}
\usepackage[english]{babel}
\usepackage[normalem]{ulem}

\newcommand{\be}{\begin{equation}}
\newcommand{\ee}{\end{equation}}
\newcommand{\bea}{\begin{eqnarray}}
\newcommand{\eea}{\end{eqnarray}}
\newcommand{\ket}{\rangle}
\newcommand{\bra}{\langle}

\newcommand{\I}{\mathds{1}}
\newcommand{\ra}{\rightarrow}

\def\C#1{\mathcal #1}

\definecolor{gray}{gray}{0.9}

\usepackage{setspace}

\begin{document}
\newtheorem{theorem}{Theorem}
\newtheorem{prop}[theorem]{Proposition}
\newtheorem{corollary}[theorem]{Corollary}
\newtheorem{open problem}[theorem]{Open Problem}
\newtheorem{conjecture}[theorem]{Conjecture}
\newtheorem{definition}{Definition}
\newtheorem{remark}{Remark}
\newtheorem{example}{Example}
\newtheorem{task}{Task}

\title{An additive refinement of quantum channel capacities}

\author{Dong-Sheng Wang}
\affiliation{Institute of Theoretical Physics, Chinese Academy of Sciences, Beijing 100190, China \\
School of Physical Sciences, University of Chinese Academy of Sciences, Beijing 100049, China}
\date{\today}
\begin{abstract}
    Capacities of quantum channels are fundamental quantities in the theory of quantum information.
    A desirable property is the additivity for a capacity measure.
    However, this cannot be achieved for a few quantities that have been established as capacity measures.  
    Asymptotic regularization is generically necessary making the study of capacities notoriously hard.
    In this work,
    by a proper refinement of the physical settings of quantum communication using restricted encodings,
    we prove additive quantities for quantum channel capacities
    that can be employed for quantum Shannon theorems.
    This refinement
    is consistent with the principle of quantum theory,
    and it further demonstrates von Neumann entropy as the cornerstone of quantum information.
\end{abstract}

\maketitle

\begin{spacing}{1}

\section{Introduction}

Entropy is an important concept in physics,
which is also the foundation of classical communication theory established by C. Shannon~\cite{Sha48}.
Information sources are modeled as random variables,
and the Shannon entropy of a source sets the bound below which 
message can be reliably transmitted with a source encoding and decoding.
Furthermore, for noisy transmission over a channel,
the maximum mutual information between the sender and receiver 
defines the channel capacity below which 
message can be reliably transmitted with a channel encoding and decoding.
Shannon's theorems guarantee the viability of error correction,
setting the foundation of information science.

The field of quantum information science has been developed for decades~\cite{NC00,KSV02,Wat18}.
The measure of quantum information is the von Neumann entropy,
and the quantum analog of Shannon's theorems and extensions have been established.
In order to harvest quantum advantages, however,
there are great challenges.
Besides the technical challenge to realize quantum error correction,
it is hard on the first hand to study the capacities of quantum channels.
Regularization is needed for both the classical and quantum capacities of a quantum channel,
i.e., a capacity may increase indefinitely when more runs of a channel are used,
while at the same time finding the optimal input achieving the capacity is not easy.
With entanglement assistance~\cite{BSS+99}, 
it turns out the capacities become additive,
hence reducing from the regularized one to the single-letter formula,
while finding the optimal assisted entangled state is also challenging.

\begin{figure}[b!]
    \centering
    \includegraphics[width=0.4\textwidth]{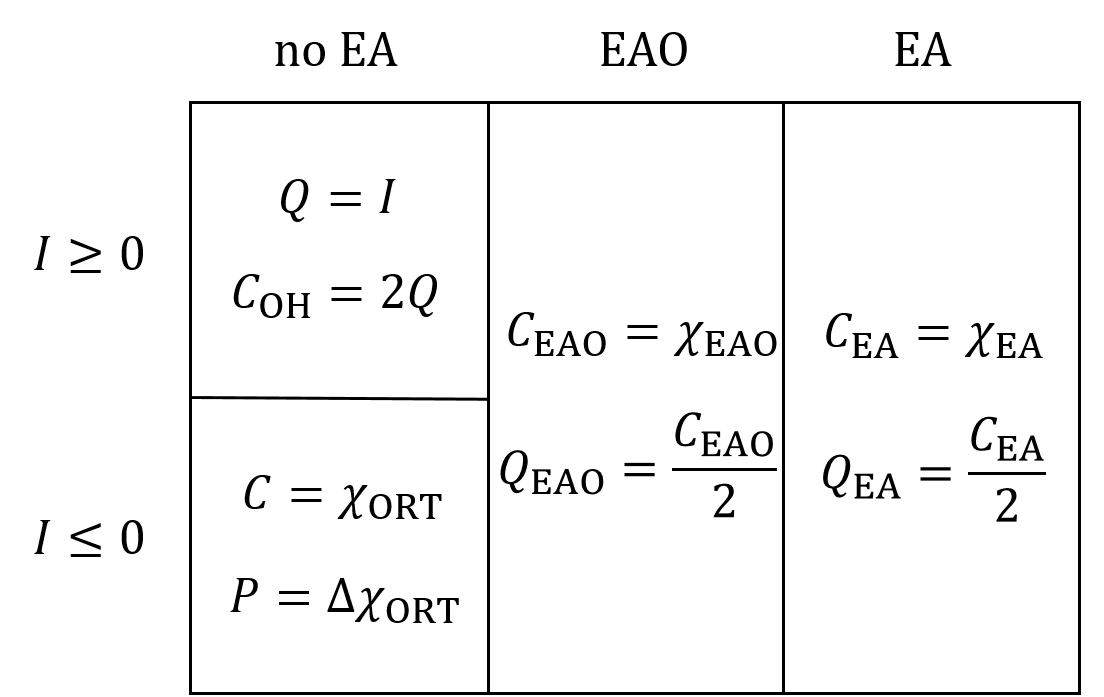}
    \caption{The capacities of quantum channels that are studied in this work which are all additive.}
    \label{fig:capacity}
\end{figure}

Compared with classical ones,
quantum states and channels obey many appealing laws,
such as the uncertainty principle and the channel-state duality~\cite{Jam72,Cho75}.
The former one has inspired great promise for quantum cryptography~\cite{BB84},
while the later one will play a central role in our study.
Due to quantum entanglement, 
quantum processes are required not only being positive,
but also completely positive, 
which brings novel structures to quantum objects such as dilation and purification.
Meanwhile, as extension of classical ones,
quantum coding has proven to be promising~\cite{LB13},
which usually does not involve heavy asymptotic analysis of channel capacities.
All these hint that quantum information is of distinct nature from classical ones,
and there could be a proper way to drop the regularization requirement on capacities of quantum channels.
In this work, we find that this can indeed be achieved 
by a slight refinement of the settings of quantum communication,
see Fig.~\ref{fig:capacity}.
What we modify is the notion of ensemble and the task for entanglement sharing
involved to define capacities.

We find that, reducing the protocol of entanglement generation~\cite{Dev05} to 
ebit distribution~\footnote{An ebit is also known as a Bell state, 
which is a bipartite maximally entangled state.},
the quantum capacity is given by the coherent information~\cite{SN96} 
of a channel when the input is the identity, hence additive. 
Using the orthogonal ensemble that we define, 
the classical capacity becomes additive
without sacrificing the quantum features,
although allowing more general ensemble and entangled input states 
can lead to peculiar phenomenon such as the super-activation~\cite{SY08}
but losing the additivity.
Our finding provides an information-theoretic interpretation of the channel-state duality,
and it confirms our intuition that,
due to the complete positivity requirement, 
it is sometimes easier to compute the quantum capacities than classical ones.
Therefore, it can simplify the proofs of quantum Shannon theorems,
e.g., the strong converse channel coding theorems~\cite{Win99,ON99,BDH+14,TWW17}. 
We also learn that quantum resources, 
such as entanglement and coherence, need to be harvested in meticulous ways,
and further study may benefit from the recently developed quantum resource theory~\cite{CG19}.
It also demonstrates the central roles of von Neumann entropy as the measure of quantum information.

\section{Quantum channels}

We study finite-dimensional Hilbert spaces.
The infinite-dimensional case needs to be treated separately.
Quantum states, or density operators $\rho \in \C D(\C X)$ 
are trace-class nonnegative semidefinite operators acting on a given Hilbert space $\C X$.
Quantum operations, also known as quantum channels,  
are described as completely positive, trace-preserving (CPTP) maps of the form
\be \Phi(\rho)= \sum_i K_i \rho K_i^\dagger, \ee
for $\Phi$ forming a convex body $C(\C X, \C Y)$, 
$\forall \rho \in \C D(\C X)$, 
and $K_i$ known as Kraus operators satisfying $\sum_i K_i^\dagger K_i= \I_d$,
for $\I_d$ denoting the identity operator of dimension $d=\text{dim}\C X$.
We will omit the subscript dimension most of the time for simplicity.

A quantum channel $\Phi$ can be dilated to an isometry $V$ with
$\Phi(\rho)=\text{tr}_\text{E} V\rho V^\dagger$ for E denoting an `environment' (or Eve).
The isometry $V$ is formed by Kraus operators $V=\sum_i |i\ket \otimes K_i$
for $|i\ket$ as orthogonal states of E. 
A complementary channel $\Phi^c$ can be defined as the map from the input to E,
and the state of E is 
\be \rho_\text{E}=\sum_{ij} \text{tr}(\rho K_i^\dagger K_j) |j\ket \bra i|. \ee

Any quantum channel $\Phi\in C(\C X, \C Y)$ can also be represented as its dual state~\cite{Jam72,Cho75}
\be \omega_{\Phi} := \Phi \otimes \I (\omega), \ee 
called a Choi state here, for $\omega:=|\omega\ket \bra \omega|$,
and $|\omega\ket:= \sum_i |ii\ket /\sqrt{d}$.
This is also known as the channel-state duality.
The rank of the channel $\text{r}(\Phi)$ is the rank of $\omega_{\Phi}$.
The state $|\omega\ket$ is a Bell state, also known as an ebit. 

Choi states are bipartite and for clarity, we label the sites as A and B.
Let $\pi_d:=\I /d$.
The partial trace of a Choi state is constrained as $\text{tr}_\text{B} \omega_\Phi= \pi_d$,
and $\text{tr}_\text{A} \omega_\Phi= \Phi(\pi_d)$.
Given $\omega_\Phi$, the action of the channel can be obtained as
\be \Phi(\rho)= d \; \text{tr}_\text{A} [\omega_\Phi (\I \otimes \rho^t) ], \label{eq:readout}\ee
for $\rho^t$ as the transpose of a state $\rho \in \C D(\C X)$.
As has been shown~\cite{W20_choi,W22_qvn}, this is sufficient to simulate the channel $\Phi$.
With a binary positive operator-valued measure (POVM) $\{\sqrt{\rho^t}, \sqrt{\I-\rho^t}\}$ acting on A,
any observation $\C O$ on $\Phi(\rho)$ can be measured on site B. 
This is heralded but deterministic, i.e., 
it belongs to the class of local operations with classical communication (LOCC).
This provides an operational interpretation of the channel-state duality.
We will use this fact to refine the notion of capacities of quantum channels.

There are a few important classes of channels we will need.
A channel is unital if $\Phi(\pi_d)=\pi_{d_o}$.
A channel is extreme iff the set $\{K_i^\dagger K_j\}$ is linearly independent~\cite{Cho75}.
A rank-two qubit channel is either extreme or unital~\cite{RSW02}.
A channel is entanglement-breaking (EB) iff its Kraus operators are all of rank one~\cite{HSR03}.
Namely, they can be written as $K_i=|x_i\ket \bra y_i|$ for un-normalized vectors $|x_i\ket$ and $|y_i\ket$.
Another equivalent condition is that the Choi state is separable.
It becomes quantum-to-classical (QC) if $\{|x_i\ket\}$ is orthonormal,
and classical-to-quantum (CQ) if $\{|y_i\ket\}$ is orthonormal.

Two central measures for distances are fidelity and trace distance.
They satisfy the Fuchs-van de Graaf inequality
\be 1-\sqrt{F(\rho,\sigma)} \leq D_t(\rho,\sigma) \leq \sqrt{1-F(\rho,\sigma)}, \label{eq:FG}\ee
$\forall \rho, \sigma \in \C D(\C X)$,
the trace distance $D_t(\rho,\sigma):=\frac{1}{2}\|\rho-\sigma\|_1$,
and the fidelity $F(\rho,\sigma):=\|\sqrt{\rho}\sqrt{\sigma}\|_1^2$,
with $\|\cdot\|_1$ denoting the trace norm.
When one state is pure, say $\sigma$, the lower bound becomes $1-F(\rho,\sigma)$.
For channels, we use the diamond-norm distance~\cite{KSV02,Wat18}
$D_\diamond(\Phi, \Psi) =\|\Phi \otimes \I- \Psi \otimes \I \|_{1\ra 1}$,
which relates to 
the trace distance between Choi states $D_t(\omega_\Phi, \omega_\Psi)$ by~\cite{Wat18}
\be \frac{1}{2d}D_\diamond( \Phi, \Psi ) \leq D_t(\omega_\Phi, \omega_\Psi) \leq \frac{1}{2} D_\diamond( \Phi, \Psi ). \label{eq:choidis} \ee
The fidelity between Choi states is known as 
the average entanglement fidelity~\cite{Sch96} 
\be F_E(\Phi, \Psi):=  F(\Phi \otimes \I (\omega),  \Psi \otimes \I (\omega)), \ee
and it holds
\be 1-\sqrt{F_E(\Phi,\Psi)} \leq D_t(\omega_{\Phi},\omega_{\Psi}) \leq \sqrt{1-F_E(\Phi,\Psi)}.\ee
If one channel is unitary, $\C U$, then
\be 1-F_E(\Phi,\C U) \leq D_t(\omega_{\Phi},\omega_{\C U}) \leq \sqrt{1-F_E(\Phi,\C U)}. \ee

\section{Entropy}

The von Neumann entropy is defined as 
\be H(\rho)=\log d-R(\rho \| \pi_d), \ee 
for the quantum relative entropy $R(\rho \| \sigma)=\text{tr} \rho \log \rho - \text{tr} \rho \log \sigma$, $\forall \rho, \sigma \in \C D(\C X)$ 
(see~\cite{Wat18} for more details).
It is monotonic $H(\Phi(\rho))\geq H(\rho)$ under unital channels $\Phi$
but not for non-unital ones.
With the eigenvalue decomposition $\rho=\sum_i p_i |i\ket \bra i|$,
a purification of it is defined as $|\varphi_\rho\ket = \sum_i \sqrt{p_i} |i\ket  |i\ket $.
The purification of $\pi_d$ is an ebit.
The coherent information $I_c(\rho,\Phi)$ is defined as~\cite{SN96,Llo97}
\be I_c(\rho,\Phi):=H(\Phi(\rho))-H(\Phi\otimes \I (\varphi_\rho) ), \ee
and we also let $J(\rho,\Phi):=H(\rho)+ I_c(\rho,\Phi)$ be called the coherent mutual information,
which is always nonnegative.
An important property is that $J(\Phi):=\max_\rho J(\rho,\Phi)$ is additive~\cite{AC97}.

The coherent information is the negation of a conditional von Neumann entropy,
and it can be negative yet with an operational interpretation called state merging~\cite{HOW05}.
It satisfies the data-processing inequality for all channels~\cite{SN96}.
Now define \be I(\Phi):=I_c(\pi_d,\Phi), \label{eq:i}\ee
which, plus $\log d$, is the coherent mutual information contained in the Choi state $\omega_\Phi$.
There exists an intriguing relation between the rank and coherent information of a channel.
Recently it is shown that~\cite{SD22} 
the maximum coherent information $I_c(\Phi):=\max_\rho I_c(\rho,\Phi)$
is almost surely positive if $\text{r}(\Phi) \leq d$; otherwise,
it is almost surely non-positive. 
It is known that EB channels, whose Choi states are separable, 
have negative or zero coherent information, and 
necessarily are of rank no smaller than $d$, 
the input system dimension.

\section{Quantum capacity}

We consider bipartite quantum communication.
Alice aims to send qubits over a noisy quantum channel to Bob with high-enough fidelity by encoding and decoding.
Given the operational meaning of Choi states (cf. Eq.~(\ref{eq:readout})),
it only needs to distribute ebits over the two parties.
With multiple copies of ebits and LOCC, Alice can send any state $\rho$ over to Bob.
Note this is a special case of general entanglement distribution task~\cite{DHW04}.
We closely follow the logic of the book~\cite{Wat18},
wherein more details can be found. 

\begin{definition}
(Ebit distribution capacity of a channel) 
Let $\Phi\in C(\C X, \C Y)$ be a channel, and an integer $m=\lfloor \alpha n \rfloor$ 
for all but finitely many positive integers $n$ and an achievable rate $\alpha \geq 0$, 
there exists channels $\C E\in C(\C Z^{\otimes m}, \C X^{\otimes n})$ 
and $\C D\in C(\C Y^{\otimes n}, \C Z^{\otimes m})$ such that the fidelity satisfies
 $F_E(\I^{\otimes m}, \C D \Phi^{\otimes n} \C E)\geq 1- \epsilon$
for every choice of a positive real number $\epsilon$,
and the ebit distribution capacity of $\Phi$, denoted $Q_{ED}(\Phi)$,
is defined as the supremum of all $\alpha$.
\end{definition}

We will first prove that $Q_{ED}$ is equal to the quantum capacity $Q$ defined below.
\begin{definition}
(Quantum capacity of a channel) 
Let $\Phi\in C(\C X, \C Y)$ be a channel, and an integer $m=\lfloor \alpha n \rfloor$ 
for all but finitely many positive integers $n$ and an achievable rate $\alpha \geq 0$, 
there exists channels $\C E\in C(\C Z^{\otimes m}, \C X^{\otimes n})$ 
and $\C D\in C(\C Y^{\otimes n}, \C Z^{\otimes m})$ such that 
$D_\diamond (\I^{\otimes m}, \C D \Phi^{\otimes n} \C E) \leq \epsilon$
for every choice of a positive real number $\epsilon$
and the quantum capacity of $\Phi$, denoted $Q(\Phi)$,
is defined as the supremum of all $\alpha$.
\end{definition}

\begin{theorem}
Let $\Phi\in C(\C X, \C Y)$ be a channel, then $Q(\Phi)=Q_{ED}(\Phi)$.
\end{theorem}
\begin{proof}
The Fuchs-van de Graaf inequality~(\ref{eq:FG}) directly yields $Q(\Phi)\leq Q_{ED}(\Phi)$.
Theorem 8.45 of~\cite{Wat18} yields $Q(\Phi)\geq Q_{ED}(\Phi)$.
\end{proof}

From the channel-state duality, it is simple to expect $Q=Q_{ED}$. 
However, the inequality between Choi states and diamond distance on channels~(\ref{eq:choidis})
is not enough to prove this due to a dimension-dependence issue. 
Theorem 8.45 of~\cite{Wat18} (based on~\cite{BKN00}) is stronger, 
which guarantees a small diamond distance via encoding and decoding when $F_E$ is large enough.
This further implies the worst-case entanglement fidelity 
$F_E'(\Phi, \Psi)= \min_{\varphi_\rho} F( \Phi \otimes \I (\varphi_\rho),  \Psi \otimes \I (\varphi_\rho))$,
for $|\varphi_\rho\ket$ as a purification of $\rho\in \C D(\C X)$, is also large enough.
This means that 
using the purification $|\varphi_\rho\ket$ to achieve its quantum capacity becomes redundant.
The property of a channel $\Phi$ is fully characterized by its Choi state $\omega_\Phi$.

The established proof of quantum coding theorem uses $I_c(\pi_d,\Phi)\leq Q_{EG}$~\cite{Wat18},
for the later as the capacity of entanglement generation~\cite{Dev05}.
Fortunately, its proof actually proves $I_c(\pi_d,\Phi)\leq Q_{ED}$,
so we can directly apply its method to prove $Q(\Phi)=I(\Phi)$.

\begin{theorem}
(Quantum capacity theorem) 
Let $\Phi\in C(\C X, \C Y)$ be a channel, then $Q(\Phi)=I(\Phi)$.
\end{theorem}
\begin{proof}
By a close examination of Theorem 8.53~\cite{Wat18}, it proves $I(\Phi)\leq Q_{ED}(\Phi)$.
So to prove $I(\Phi)\geq Q_{ED}(\Phi)$, 
let $\alpha$ and $\delta$ satisfy $Q_{ED}(\Phi)$, then 
\be H(\C D \Phi^{\otimes n} \C E \otimes \I^{\otimes m} (\omega^{\otimes m}))\leq 2\delta m+1, \ee
and 
\be H(\C D \Phi^{\otimes n} \C E (\I^{\otimes m}))\geq m-\delta m-1, \ee
so \be I(\Phi) \geq (1-3\delta) \alpha -\frac{3}{n}, \ee
which implies $I(\Phi)\geq Q(\Phi)=Q_{ED}(\Phi)$.
\end{proof}
Compared with the original quantum capacity formula $\lim_{n\ra \infty}\frac{I_c(\Phi^{\otimes n})}{n}$,
this is a huge reduction. 
It is due to two facts: the additivity of $I(\Phi)$ defined in Eq.~(\ref{eq:i}) and the ED protocol.
We may name it the ``refined quantum capacity'';
however, for simplicity, we call it the quantum capacity in this work.
It was established that 
the quantum capacity $Q$ is equal to entanglement generation capacity $Q_{EG}$~\cite{Dev05}.
A careful look reveals that such an equivalence is relative to the set of entangled states
as assistance or resource.
On the contrary, there is no such resource requirement for ED,
or in other words, the input to the channel for ED is limited. 
We find that a more precise formulation would require quantum resource theory, 
which is left for a separate study~\cite{WLWL24}. 
Also note the ED protocol is not the same as the 
entanglement-assisted (EA) setting studied later on.

The $I(\Phi)$ is additive and bounded as $|I(\Phi)| \leq \log d$.
It is invariant 
$I(\Phi)=I(\C U \circ \Phi \circ \C V)$ for any isometry $\C U$ and $\C V$.
The values of it for some familiar channels are shown in Table~\ref{tab:cap-example}.

\begin{table}[]
    \centering
    \begin{tabular}{||c|c||} \hline
    channel & quantum capacity  \\ \hline 
    replacement &  $-\log d$ \\ \hline 
    isometry & $\log d$ \\ \hline 
    AD ($d=2$) & $h(\frac{1+\gamma}{2})-h(\frac{\gamma}{2})$ \\ \hline  
    unital & $\log d - H(p_i)$ \\ \hline  
    erasure & $(1-2p)\log d$ \\ \hline
    dephasing & $\geq 0$ \\ \hline 
    EB & $\leq 0$ \\ \hline 
    \end{tabular}
    \caption{The refined quantum capacities of a few channels.
    Details are found in the appendix.}
    \label{tab:cap-example}
\end{table}

By quantum teleportation (QT)~\cite{BBC+93} and superdense coding (DC)~\cite{BW92},
we can treat a qubit as two bits: one bit for phase information (for Pauli $Z$),
and one bit for bit-value information (for Pauli $X$).
Given an ebit, we can send two bits by just sending one qubit via DC,
or send one qubit by just sending two bits via QT.
So for $Q\geq 0$, we can define a coherent classical capacity $C_\textsc{oh}=2Q$.
Note this is different from the classical capacity based on Holevo quantity studied below.

\begin{corollary}
(Coherent classical capacity) Let $\Phi\in C(\C X, \C Y)$ be a channel.
If $Q(\Phi) \geq 0$, 
the coherent classical capacity is $C_\textsc{oh}(\Phi)=2Q(\Phi)$.
\end{corollary}

The quantum converse coding theorem proves that if the information rate $R> Q(\Phi)$,
then it is not possible to reliably transmit qubits. 
With the refined capacity, 
it is clear to see the strong converse coding theorem holds via the 
quantum Gallager’s exponent formula~\cite{SW13}.

If $Q<0$, we can define a classical capacity $C$, which can only send commuting states,
and many others.
We cannot define an incoherent quantum capacity as $C/2$ since a quantum state
is not a mixture of an $X$ part and an $Z$ part. 
This signifies the uniqueness of quantum entanglement,
which coherently combines the bit and phase information. 

\section{Classical capacity}

To transmit classical bits over quantum channels,
Alice will use a quantum ensemble $\{p_i,\rho_i\}$ obtained by
a classical-to-quantum encoding $x_i \mapsto \rho_i$ to represent a
random variable $\text{X}=\{p_i, x_i\}$,
and Bob will do measurement and obtain a random variable Z~\footnote{Note here X and Z are merely symbols for classical random variables instead of Pauli operators.}.
For this setting, we find that the quantum ensemble is required as resources,
which, however, may be unnecessary.
Therefore, in a similar spirit with the quantum case,
we will require the encoding to be isometric so that $\{\rho_i\}$ is an orthogonal set.
However, this is the most succinct setting,
and it can also be extended properly, e.g., by allowing approximations~\cite{WWC+22}.

\begin{definition}
(An ``orthogonal'' (ORT) ensemble) Let $\{\rho_i\}$ be a set of states with orthogonal supports,
 then $\{p_i, \rho_i\}$ is an ORT ensemble.
\end{definition}
We will restrict to ORT ensemble of pure states.
For a random variable $\text{X}=\{p_i, x_i\}$, a quantum encoding $|x_i\ket$ of $x_i$,
and a channel $\Phi \in C(\C X, \C Y)$ we define 
\be \chi_\textsc{ort}(p_i,\Phi):=H(\sum_i p_i \Phi(|x_i\ket \bra x_i|))-\sum_i p_i H(\Phi(|x_i\ket \bra x_i|)), \ee
which is the mutual information in the classical-quantum state
\be \rho_{XY}=\sum_i p_i |x_i\ket \bra x_i| \otimes \Phi(|x_i\ket \bra x_i|), \ee
and $\chi_\textsc{ort}(\Phi):=\max_{p_i} \chi_\textsc{ort}(p_i,\Phi)$.
It is clear to see $\chi_\textsc{ort}(\Phi)$ is additive
while the standard Holevo quantity $\chi(\Phi)$ is 
known not to be, in general,
as the later allows more general ensembles.

Now we can refine the notable Holevo-Schumacher-Westmoreland theorem 
which established the classical capacity of a quantum channel
to be $\lim_{n\ra \infty}\frac{\chi(\Phi^{\otimes n})}{n}$.
\begin{theorem}
(Classical capacity theorem) 
Let $\Phi\in C(\C X, \C Y)$ be a channel, then $C(\Phi)=\chi_\textsc{ort}(\Phi)$.
\end{theorem}
\begin{proof}
As it is known $\chi(\Phi) \leq C(\Phi)$~\cite{Wat18} and $\chi_\textsc{ort}(\Phi)\leq \chi(\Phi)$,
so $\chi_\textsc{ort}(\Phi) \leq C(\Phi)$.
To prove $\chi_\textsc{ort}(\Phi) \geq C(\Phi)$, we use the same method
as that of Theorem 8.27~\cite{Wat18}
but only with incoherent ensemble $\{\frac{1}{2^m},{\bf b}_m\}$,
for bit-string ${\bf b}_m:=b_1\cdots b_m$,
and find the classical mutual information $I(\text{X}:\text{Z}) \leq n \chi_\textsc{ort}(\Phi)$,
which leads to $\chi_\textsc{ort}(\Phi) \geq C(\Phi)$.
\end{proof}
This reduction is due to the additivity of $\chi_\textsc{ort}(\Phi)$ and the usage of
orthogonal ensemble.
This is physically reasonable since
it is natural to treat a bit-string $x_i$
as a bit-string state $|x_i\ket$,
and non-isometric encoding itself
will blur the message on the first hand.
In other words, there is no entanglement in $|x_i\ket$
and the state entering $\Phi$ is pure,
but the decoding scheme is entangling as the standard setting. 
A local setting of the classical capacity is also discussed in Chapter 20 of the book~\cite{Wil17},
which yet allows non-isometric encoding but only local decoding.

Same as the classical case,
the optimization over $p_i$ is necessary since now
the channels do not need to preserve entanglement and the channel-state duality shall not apply.
For the set of channels that cannot distribute ebits,
they are yet at least as powerful as classical channels.
For any classical channel modeled as a stochastic map which is positive instead of completely positive,
it can be simulated by a POVM $\{F_j\}$ with 
the stochastic matrix as $(\bra i|F_j|i\ket)_{ij}$. 
An input random variable is a diagonal state 
$\rho=\sum_i p_i |i\ket \bra i|$,
and the channel is simulated by a quantum-classical channel 
\be \Phi(\rho)=\sum_j \text{tr}(F_j \rho) |j\ket \bra j|. \ee 
If the POVM is projective, then it simulates a doubly-stochastic map.
This also shows that EB channels can perform certain tasks
that may have an advantage over classical channels,
which we do not pursue here.

\section{Private capacity}

Based on $\chi_\textsc{ort}(\Phi)$, it is straightforward 
to obtain an additive private capacity. 
Let $\Delta\chi_\textsc{ort}(p_i, \Phi):=\chi_\textsc{ort}(p_i, \Phi)-\chi_\textsc{ort}(p_i, \Phi^c)$,
then we define an additive private capacity
\be P(\Phi):=\max_{p_i} \Delta\chi_\textsc{ort}(p_i, \Phi), \ee
for $\Phi^c$ as the complementary channel of $\Phi$.
It holds 
\be 0\leq P(\Phi)\leq C(\Phi). \ee

Indeed, privacy is different from entanglement,
although entanglement is already a kind of privacy due to the uncertainty principle.
There are channels with zero quantum capacity but an arbitrarily large private capacity~\cite{LLS+14}.
Under the completely dephasing channel $\Delta$, 
the ebit becomes the maximally classically-correlated state,
a ``debit.''
The key difference between them is that ebits allow the transmission of qubits.

\section{Entanglement-assisted setting}

It is well established that using entanglement assistance 
can lead to additive settings~\cite{BSS+99}.
This is actually a great example of quantum resource theory~\cite{CG19}.
In this section, we find that there is also a refinement for the entanglement-assisted (EA) setting. 
We first refine the EA setting to only use ebits as the assistance
for encoding and decoding,
and then compare with the standard EA setting.

\begin{definition}
(``EA orthogonal'' (EAO) ensemble) Let $\{U_i\}$ be a set of orthogonal unitary operators.
Let $|\omega_{U_i}\ket=U_i\otimes \I |\omega\ket$, then $\eta:=\{p_i, |\omega_{U_i}\ket\}$ is an EAO ensemble.
\end{definition}
We now introduce
\be \chi_\textsc{eao}(\eta,\Phi):= H(\sum_i p_i \Phi\otimes \I(\omega_{U_i}))-\sum_i p_i H(\Phi\otimes \I(\omega_{U_i})), \ee
and $\chi_\textsc{eao}(\Phi):=\max_{p_i}\chi_\textsc{eao}(\eta,\Phi)$.
This can be viewed as encoding $x_i$ by a set of orthogonal unitary operators $U_i$.
It is clear that $\chi_\textsc{eao}(\Phi)$ is additive.
Recall that an EA ensemble can be more general:
the ebit $|\omega\ket$ can be replaced by any bipartite pure state,
and the set of $U_i$ replaced by a set of channels~\cite{Wat18}. 

With the same method for proving $\chi_\textsc{ort}(\Phi)$ while using EAO ensemble, it is clear to prove:
\begin{theorem}
(EAO-classical capacity theorem) 
Let $\Phi\in C(\C X, \C Y)$ be a channel, then $C_\textsc{eao}(\Phi)=\chi_\textsc{eao}(\Phi)$.
\end{theorem}
Although $\chi_\textsc{eao}(\Phi)$ also involves an optimization over $p_i$,
it turns out this is unnecessary due to an intriguing relation between
$\chi_\textsc{ea}(\Phi)$ and coherent mutual information.
Note we use $\chi_\textsc{ea}$ to denote the usual EA Holevo quantity $\chi_E$ in this paper~\cite{Wat18}.

\begin{corollary}
Let $\Phi\in C(\C X, \C Y)$ be a channel, then $C_\textsc{eao}(\Phi)=\chi_\textsc{eao}(\Phi)=\log d+I(\Phi)$.
\end{corollary}
\begin{proof}
In Lemma 8.39 of Ref.~\cite{Wat18}, it proves 
\be \chi_\textsc{ea}(\Phi)\geq H(\pi_d)+I_c(\pi_d, \Phi) \ee
for $\pi_d=\Pi/d$ and $d=\text{tr}(\Pi)$ and $\Pi$ is a projector.
The proof actually proves the stronger result
\be \chi_\textsc{eao}(\Phi)\geq H(\pi_d)+I_c(\pi_d, \Phi) \ee
as it uses a completely uniform ensemble of Bell states, $\eta_*$, which is an EAO ensemble.
Also it proves 
\be \chi_\textsc{eao}(\eta_*, \Phi)= \log  d+I(\Phi), \ee
which is the equality between two mutual information quantities.

The Lemma 8.40~\cite{Wat18} also applies to any EAO ensemble $\eta_\textsc{eao}$ which becomes 
\be \chi_\textsc{ort}(\Phi\otimes \I (\eta_\textsc{eao}))\leq \log d +I(\Phi). \ee
This completes the proof.
\end{proof}

By QT and DC, we define \be Q_\textsc{eao}(\Phi):=(\log d+ I(\Phi))/2. \ee
This means that when $I(\Phi)<0$, yet with ebits we can recover its ability to transmit qubits
so that the whole channel ($\Phi$ assisted by ebits) is still quantum.
The information is still directly transmitted from $\Phi$ so can only be classical,
therefore, $\log d+I(\Phi)$ is an EAO classical capacity instead of quantum capacity.
It is clear to show $Q(\Phi)=I(\Phi)\leq Q_\textsc{eao}(\Phi)$ since $\log d\geq I(\Phi)$. 

Actually, given $\Phi$, the optimal scheme to achieve $C_\textsc{eao}(\Phi)$ is the superdense coding, 
which encodes bits into Pauli operators hence the set of ebits.
The decoding is formed by the decoder of $\Phi$ followed by Bell measurements.
Also, modular the ebit, a qubit is equivalent to two bits.
As QT converts a qubit into two bits, 
so it can be viewed as an EAO scheme with $\Phi$ acting on the bit communication channel.
That is, QT is optimal to achieve $Q_\textsc{eao}(\Phi)$.

The standard EA setting uses more general entangled states as the assistance.
Importantly, the maximum coherent mutual information $J(\Phi)$ is additive.
There exists a gap between $\log d+ I(\Phi)$ and $J(\Phi)$ for some channel $\Phi$.
Achieving the maximum of $J(\rho,\Phi)$ by a state $\rho$
is to achieve the regularized EA Holevo capacity,
which shall require a highly-entangled state $\xi$ as a free assistance which is not a collection of ebits.
It is not easy to prepare and distribute such states, however.
This is similar with the situation for entanglement generation which also 
assumes the free assistance of entangled states~\cite{WLWL24}.
For our reduced case, we do not take them to be free assistance or resources.

Given a channel, to see if there is a gap between the EA and EAO settings is interesting but not easy.
Here we consider the class of CQ channels, which have negative quantum capacity.
For a CQ channel acting on the purification of a state $\rho=\sum_k p_k |k\ket \bra k|$,
the mutual information is 
\be J(\rho,\Phi)=H(\sum_k p_k R_k)-\sum_k p_k H(R_k), \ee
for a set of states $\{R_k\}$.
The EA classical capacity is $C_\textsc{ea}(\Phi)=\max_\rho J(\rho,\Phi)$ which is to optimize over $p_k$.
If $R_k:=\Psi(|k\ket\bra k|)$, then it is easy to see $C_\textsc{ea}(\Phi)$ is the same as the classical capacity of $\Psi$
which is $\chi_\textsc{ort}(\Psi)$.
Indeed, to achieve optimal classical capacity, 
the optimal input distribution is not always the uniform one.
Hence using $\varphi_\rho$ as a resource for EA setting is reasonable in this case.
For QC channels, the $\chi_\textsc{eao}$ is $\log d$, 
so it is the optimal $\chi_\textsc{ea}$.
For channels with positive quantum capacity,
it is not clear how to find the optimal resource entangled states.

\section{Conclusion}

To conclude, we refined the settings of quantum communication in order to obtain additive capacities of quantum channels.
Surprisingly, it is straightforward to achieve additivity and the proof follows almost exactly the same as the original ones~\cite{Wat18}.
It also highlights the role of channel-state duality,
which indeed is a unique feature of quantum information.

Now an immediate question arises: if the refined additive case 
works, then how about the standard ``non-additive world''? 
A primary comparison reveals that the general quantum resource theory~\cite{CG19} 
should be employed to characterize hierarchy of quantum resources in various 
quantum communication settings~\cite{WLWL24},
as has been done recently for universal quantum computing models~\cite{W23_ur}.
This, together with other important directions such as 
the infinite-dimensional case~\cite{HW01,WPG+12},
detailed proofs of quantum Shannon theorems~\cite{Win99,ON99,BDH+14,TWW17},
and application in quantum error correction and fault-tolerance~\cite{NC00},
shall be investigated further.

Besides, Renyi entropies are also well defined and widely used~\cite{Wat18}, 
as extensions of Shannon and von Neumann entropy.
It is clear that the classical and private capacities 
can be extended to Renyi entropies.
However, the proofs for quantum and EA capacities rely on the strong subadditivity of the von Neumann entropy,
which is not satisfied by Renyi entropies in general, 
so cannot be extended.
However, it remains to see if quantum Renyi entropies of special forms 
can be used to define quantum and EA capacities.

\section{Acknowledgement}
This work has been funded by
the National Natural Science Foundation of China under Grants
12047503 and 12105343.
Comments from Y.-D. Liu, Y.-J. Wang, 
M. Hayashi, S.-L. Luo, J. Watrous, H. Zhu,
and an anonymous referee are acknowledged.

\appendix 

\section{Examples}

Here we compute the refined quantum capacity of a few types of quantum channels.

For a $d$-dimensional input system,
the replacement channels on it have minimal quantum capacity $-\log d$, 
whose Choi states are product states. 
The unitary and isometry have maximal quantum capacity 
$\log d$, whose Choi states are maximally entangled states. 
An erasure channel is often defined as 
\be \Phi(\rho)= (1-p)\rho +p |e\ket \bra e| \ee
for $\bra e|\rho|e\ket=0$, $\forall \rho\in \C D(\C X)$.
It is easy to find $I(\Phi)=(1-2p)\log d$. 
This agrees with the known result~\cite{BDS97}. 

\subsection{Unital channels}

A large class is unital channels that preserves identity. 
Let $\{K_i\}$ be the canonical form of a unital channel $\Phi \in C(\C X, \C Y)$, then 
\be I(\Phi)= \log d_o -H(p_i) \ee
for $p_i=\text{tr}(K_i^\dagger K_i)/d$,
and $d=\text{dim}\C X$, $d_o=\text{dim}\C Y$.
It achieves the maximal value of $\log d$ for unitary and isometry,
and minimal value of $-\log d$ for rank-$(dd_o)$ channels with $p_i=1/(dd_o)$.

It includes Pauli, depolarizing, dephasing channels as special cases. 
For a Pauli channel, the Kraus operators are orthogonal Pauli operators $\sigma_i$ with different probabilities. 
The depolarizing channels are special types of Pauli channels 
with equal probability for nontrivial Pauli operators.
Let a depolarizing channel be 
\be \Phi(\rho)=(1-p)\rho+ \frac{p}{d^2-1}\sum_{i=1}^{d^2-1}\sigma_i(\rho), \ee
and its quantum capacity $I(\Phi)$ can be negative. 
We find the value of $p$ at which $I(\Phi)=0$ is a monotonically increasing function of $d$,
from $\approx 0.19$ at $d=2$ to 0.5 for $d\ra \infty$.

A dephasing channel can be viewed as a Pauli channel with only $Z$-type Pauli operators.
A dephasing channel can also be treated as a Schur-product channel $\rho \mapsto \rho \circ S$
for a correlation matrix $S$.
Let $S:=\sum_{ij}\bra \psi_j|\psi_i\ket |i\ket \bra j|$ for $|\psi_i\ket=U_i|0\ket$ 
with a set of unitary operators $\{U_i\}$, 
then the Kraus operators are $K_j=\sum_{i=1}^d \bra j|\psi_i\ket |i\ket \bra i|$.
The quantum capacity is $I(\Phi)=\log d - H(\rho_E)$ for $\rho_E=\sum_i |\psi_i\ket
\bra \psi_i|/d$.
It reduces to the Pauli description if $\{U_i\}$ are from Pauli $Z$-type operators.
It is always non-negative since the rank is upper bounded by $d$, and
it is zero if $\rho_E=\pi_d$.

\begin{figure}[t!]
    \centering
    \includegraphics[width=0.5\textwidth]{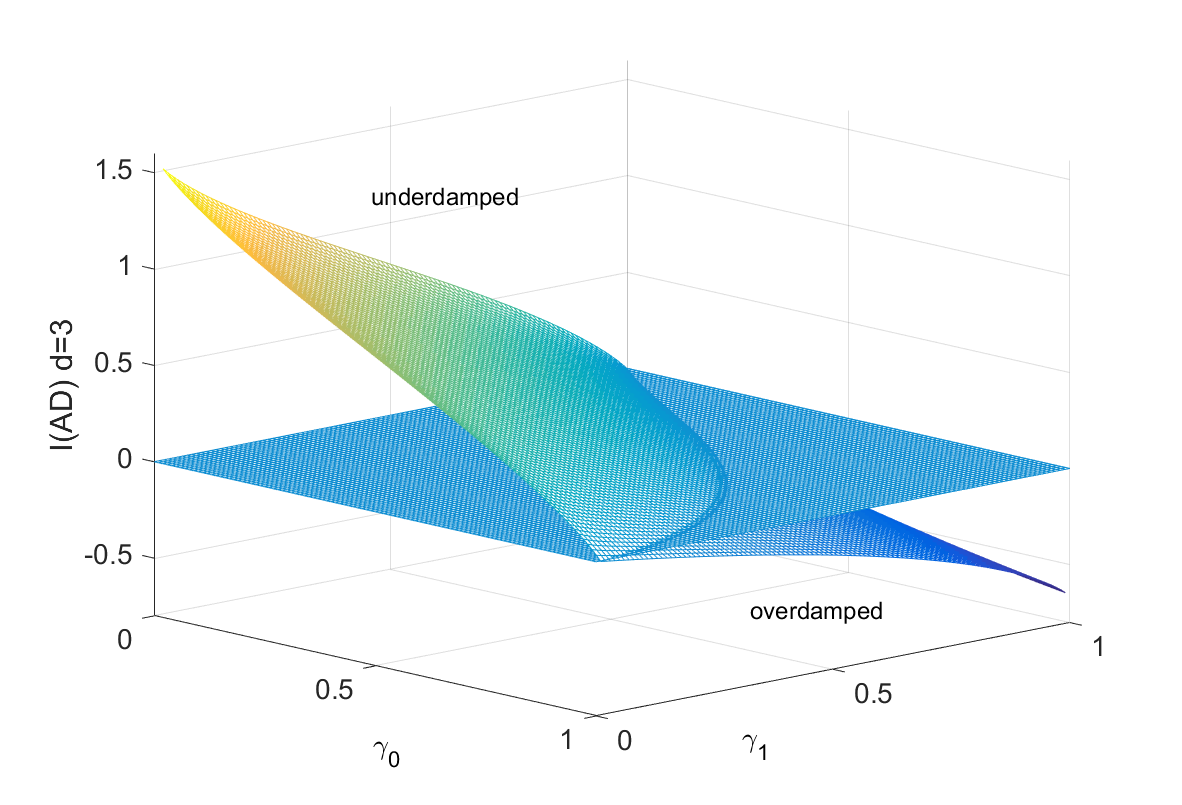}
    \caption{The capacity of the qutrit amplitude-damping channel.}
    \label{fig:capacity3}
\end{figure}

\subsection{Entanglement-breaking channels}

The Holevo form of an EB channel is defined by a POVM $\{E_b\}$ and a collection of states $\{R_a\}$ so that 
\be \Phi(\rho)=\sum_{ab} R_a \text{tr}(E_b \rho), \ee 
and its Choi state $\omega_\Phi=\frac{1}{d}\sum_{ab}R_a\otimes E_b^t$ is separable.
For separable states, 
it is established that the local entropy is smaller than global entropy 
$S_A, S_B \leq S_{AB}$~\cite{HHH+09}.
Therefore, it is clear that $I(\Phi)\leq 0$.
For a different representation, which uses rank-one Kraus operators 
$K_i=|x_i\ket \bra y_i|$,
it turns out 
\be \Phi^c(\rho)= X \circ Y_\rho, \ee 
for $X=(\bra x_i|x_j\ket)_{ij}$, $Y_\rho=(\bra y_i|\rho|y_j\ket)_{ij}$,
which can be viewed as an extended dephasing channel~\cite{KMN+07}.
So the quantum capacity of the extended dephasing channel is always nonnegative.

An EB channel is quantum-to-classical (QC) if $\{|x_i\ket\}$ is orthonormal,
or classical-to-quantum (CQ) if $\{|y_i\ket\}$ is orthonormal.
The quantum capacity of QC channels is zero
since they convert qubits to bits.
While CQ channels convert bits to qubits, which have local coherence, 
they break entanglement.
This shows the difference between entanglement and coherence.
For CQ channels, it is the local coherence (non-orthogonal states) that makes the capacity negative. 
Indeed, the local non-orthogonality is a kind of obstacle to create entanglement,
which intends to suppress local coherence while enhance nonlocal coherence.

\begin{figure}[t!]
    \centering
    \includegraphics[width=0.5\textwidth]{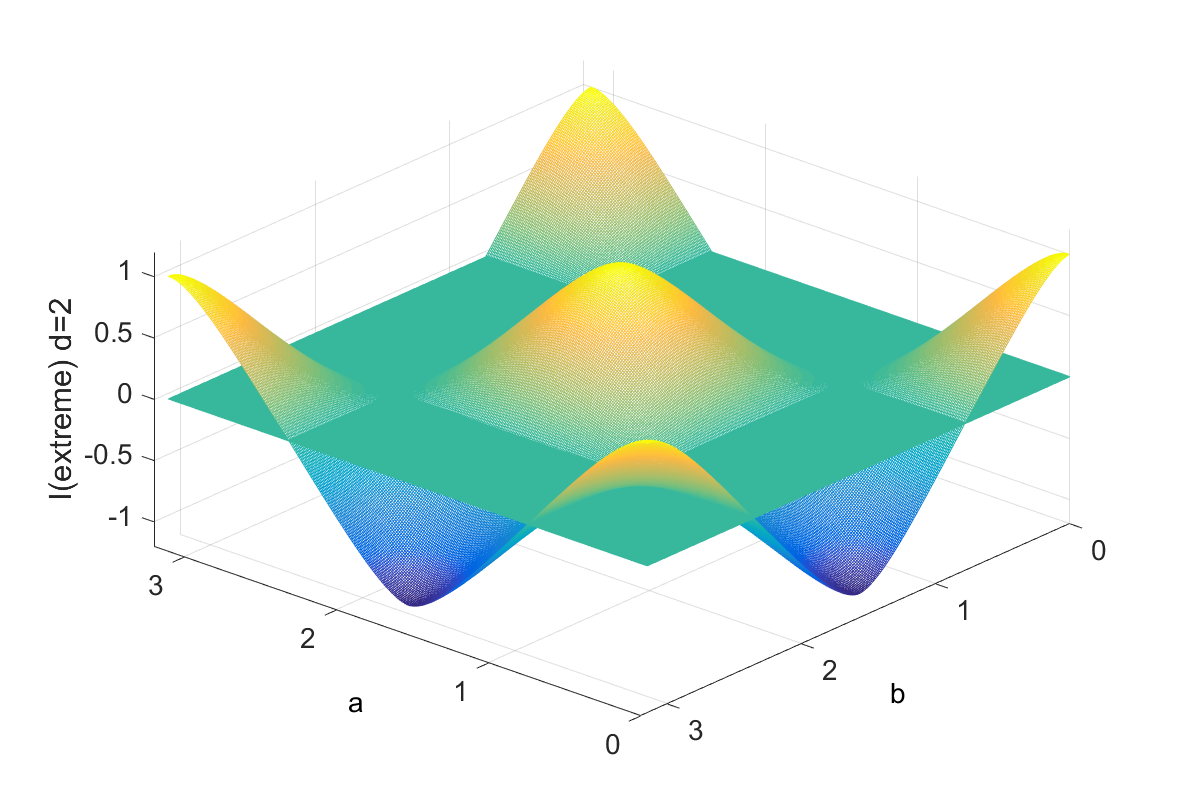}
    \caption{The capacity of the qubit extreme channel.}
    \label{fig:capacity2}
\end{figure}

\subsection{Amplitude-damping channels}

The qubit amplitude-damping (AD) channel is an extreme channel.
It can be extended to the qudit case~\cite{CG21} but not unique.
Here we define a type of qudit AD channel with Kraus operators
\be K_n= \sqrt{\gamma_n} |n\ket \bra n+1 |, \; n=0,1,\dots, d-2, \ee
and $K_{d-1}=|0\ket \bra 0| +\sum_{n=0}^{d-2} \sqrt{1-\gamma_n} |n+1\ket \bra n+1 |$,
$\gamma_n\in [0,1]$,
which is extreme in the set of qudit channels.
It can be realized by a quantum circuit of a sequence of system controlled-$R_y(\theta)$,
followed by ancilla-controlled-nots~\cite{NC00}. 

We find its quantum capacity is
\be I(\Phi)=H(\vec{p})-H(\vec{q}) \ee
for 
$\vec{p}=(1+\gamma_0, 1+\gamma_1-\gamma_0, \dots, 1+\gamma_{d-2}-\gamma_{d-3}, 1-\gamma_{d-2})/d$,
and $\vec{q}=(\gamma_0,\gamma_1,\dots,\gamma_{d-2},d-\sum_{n=0}^{d-2} \gamma_n)/d$.
It can be either positive or negative.

For the qubit case, we find 
\be I(\Phi)=h(\frac{1+\gamma}{2})-h(\frac{\gamma}{2}), \ee which ranges from 1 to -1, and equals 0 for $\gamma=0.5$.
Note $h$ is the binary entropy.
The region of $\gamma\in [0,0.5]$ is under-damped,
and $\gamma\in [0.5,1]$ is over-damped.

For the qutrit case, we find 
\bea \nonumber
I(\Phi) &=&H(\frac{1+\gamma_0}{3},\frac{1+\gamma_1-\gamma_0}{3},\frac{1-\gamma_1}{3}) \\ &-&
H(\frac{\gamma_0}{3},\frac{\gamma_1}{3},\frac{3-\gamma_0-\gamma_1}{3}). \eea 
There is also an under-damped and over-damped region, see Fig.~\ref{fig:capacity3},
while a critical line for the zero capacity.

The AD channel is a type of non-unital extreme channel.
For the qubit case, for Kraus operators 
$K_0=c_a|0\ket \bra 0|+c_b|1\ket \bra 1|$,
$K_1=s_a|1\ket \bra 0|+s_b|0\ket \bra 1|$,
with $c$ ($s$) denoting $\cos$ ($\sin$) and $a, b$ are angles, 
we find 
\be I(\Phi)=h(\frac{c_a^2+s_b^2}{2})-
h(\frac{s_a^2+s_b^2}{2}), \ee 
shown in Fig.~\ref{fig:capacity2},
which is the same as, but easier to compute than
the previous result found by optimization~\cite{WP07}.

\end{spacing}
\bibliography{ext}{}

\begin{thebibliography}{10}
\expandafter\ifx\csname url\endcsname\relax
  \def\url#1{\texttt{#1}}\fi
\expandafter\ifx\csname urlprefix\endcsname\relax\def\urlprefix{URL }\fi
\expandafter\ifx\csname href\endcsname\relax
  \def\href#1#2{#2} \def\path#1{#1}\fi

\bibitem{Sha48}
C.~Shannon, A mathematical theory of communication, {The Bell System Technical
  Journal} 27 (1948) 379.

\bibitem{NC00}
M.~A. Nielsen, I.~L. Chuang, Quantum Computation and Quantum Information,
  Cambridge University Press, Cambridge U.K., 2000.

\bibitem{KSV02}
A.~Kitaev, A.~H. Shen, M.~N. Vyalyi, Classical and Quantum Computation, Vol.~47
  of Graduate Studies in Mathematics, American Mathematical Society,
  Providence, 2002.

\bibitem{Wat18}
J.~Watrous, {The Theory of Quantum Information}, Cambridge University Press,
  2018.

\bibitem{BSS+99}
C.~H. Bennett, P.~W. Shor, J.~A. Smolin, A.~V. Thapliyal, Entanglement-assisted
  classical capacity of noisy quantum channels, Phys. Rev. Lett. 83 (1999)
  3081--3084.

\bibitem{Jam72}
A.~Jamio{\l}kowski, Linear transformations which preserve trace and positive
  semidefiniteness of operators, Rep. Math. Phys. 3 (1972) 275.

\bibitem{Cho75}
M.-D. Choi, Completely positive linear maps on complex matrices, Linear Algebra
  Appl. 10 (1975) 285--290.

\bibitem{BB84}
C.~H. Bennett, G.~Brassard, Quantum cryptography: Public key distribution and
  coin tossing, in: Proceedings of the IEEE International Conference on
  Computers, Systems and Signal Processing, Bangalore, India, (IEEE, New York),
  1984, p. 175–179.

\bibitem{LB13}
D.~Lidar, T.~A. Brun (Eds.), Quantum error correction, Cambridge University
  Press, 2013.

\bibitem{Dev05}
I.~Devetak, {The private classical capacity and quantum capacity of a quantum
  channel}, IEEE Trans. Inf. 51~(1) (2005) 44--55.

\bibitem{Note1}
An ebit is also known as a Bell state, which is a bipartite maximally entangled
  state.

\bibitem{SN96}
B.~Schumacher, M.~A. Nielsen, Quantum data processing and error correction,
  Phys. Rev. A 54 (1996) 2629--2635.

\bibitem{SY08}
G.~Smith, J.~Yard, Quantum communication with zero-capacity channels, Science
  321 (2008) 1812.

\bibitem{Win99}
A.~Winter, Coding theorem and strong converse for quantum channels, IEEE Trans.
  Inf. Theory 45 (1999) 2481.

\bibitem{ON99}
T.~Ogawa, H.~Nagaoka, Strong converse to the quantum channel coding theorem,
  IEEE Trans. Inf. Theory 45 (1999) 2486.

\bibitem{BDH+14}
C.~H. Bennett, I.~Devetak, A.~W. Harrow, P.~W. Shor, A.~Winter, The quantum
  reverse shannon theorem and resource tradeoffs for simulating quantum
  channels, IEEE Trans. Inf. Theory 60 (2014) 2926.

\bibitem{TWW17}
M.~Tomamichel, M.~M. Wilde, A.~Winter, Strong converse rates for quantum
  communication, IEEE Trans. Inf. Theory 63 (2017) 715.

\bibitem{CG19}
E.~Chitambar, G.~Gour, Quantum resource theories, Rev. Mod. Phys. 91 (2019)
  025001.

\bibitem{W20_choi}
D.-S. Wang, Choi states, symmetry-based quantum gate teleportation, and
  stored-program quantum computing, Phys. Rev. A 101 (2020) 052311.

\bibitem{W22_qvn}
D.-S. Wang, {A prototype of quantum von Neumann architecture}, Commun. Theor.
  Phys. 74 (2022) 095103.

\bibitem{RSW02}
M.~B. Ruskai, S.~Szarek, E.~Werner, An analysis of completely-positive
  trace-preserving maps on m2, Linear Algebra Appl. 347~(1--3) (2002) 159--187.

\bibitem{HSR03}
M.~Horodecki, P.~Shor, M.~B. Ruskai, Entanglement breaking channels, Rev. Math.
  Phys. 15 (2003) 629--641.

\bibitem{Sch96}
B.~Schumacher, Sending entanglement through noisy quantum channels, Phys. Rev.
  A 54 (1996) 2614--2628.

\bibitem{Llo97}
S.~Lloyd, Capacity of the noisy quantum channel, Phys. Rev. A 55 (1997)
  1613--1622.

\bibitem{AC97}
C.~Adami, N.~J. Cerf, {von Neumann capacity of noisy quantum channels}, Phys.
  Rev. A 56 (1997) 3470--3483.

\bibitem{HOW05}
M.~Horodecki, J.~Oppenheim, A.~Winter, Partial quantum information, Nature 436
  (2005) 673–676.

\bibitem{SD22}
S.~Singh, N.~Datta, {Coherent information of a quantum channel or its
  complement is generically positive}, Quantum 6 (2022) 775.

\bibitem{DHW04}
I.~Devetak, A.~W. Harrow, A.~Winter, A family of quantum protocols, Phys. Rev.
  Lett. 93 (2004) 230504.

\bibitem{BKN00}
H.~Barnum, E.~Knill, M.~Nielsen, {On quantum fidelities and channel
  capacities}, IEEE Trans. Inf. 46~(4) (2000) 1317--1329.

\bibitem{WLWL24}
D.-S. Wang, Y.-D. Liu, Y.-J. Wang, S.~Luo, Quantum resource theory of coding
  for error correction, Phys. Rev. A 110 (2024) 032413.

\bibitem{BBC+93}
C.~H. Bennett, G.~Brassard, C.~Cr\'epeau, R.~Jozsa, A.~Peres, W.~K. Wootters,
  Teleporting an unknown quantum state via dual classical and
  einstein-podolsky-rosen channels, Phys. Rev. Lett. 70 (1993) 1895--1899.

\bibitem{BW92}
C.~H. Bennett, S.~J. Wiesner, {Communication via one- and two-particle
  operators on Einstein-Podolsky-Rosen states}, Phys. Rev. Lett. 69 (1992)
  2881--2884.

\bibitem{SW13}
N.~Sharma, N.~A. Warsi, Fundamental bound on the reliability of quantum
  information transmission, Phys. Rev. Lett. 110 (2013) 080501.

\bibitem{Note2}
Note here X and Z are merely symbols for classical random variables instead of
  Pauli operators.

\bibitem{WWC+22}
D.-S. Wang, Y.-J. Wang, N.~Cao, B.~Zeng, R.~Laflamme, Theory of quasi-exact
  fault-tolerant quantum computing and valence-bond-solid codes, New J. Phys.
  24 (2022) 023019.

\bibitem{Wil17}
M.~Wilde, Quantum Information Theory, Cambridge University Press, 2017.

\bibitem{LLS+14}
D.~Leung, K.~Li, G.~Smith, J.~A. Smolin, Maximal privacy without coherence,
  Phys. Rev. Lett. 113 (2014) 030502.

\bibitem{W23_ur}
D.-S. Wang, {Universal resources for quantum computing}, Commun. Theor. Phys.
  75 (2023) 125101.

\bibitem{HW01}
A.~S. Holevo, R.~F. Werner, Evaluating capacities of bosonic {G}aussian
  channels, Phys. Rev. A 63 (2001) 032312.

\bibitem{WPG+12}
C.~Weedbrook, S.~Pirandola, R.~Garc\'{\i}a-Patr\'on, N.~J. Cerf, T.~C. Ralph,
  J.~H. Shapiro, S.~Lloyd, Gaussian quantum information, Rev. Mod. Phys. 84
  (2012) 621--669.

\bibitem{BDS97}
C.~H. Bennett, D.~P. DiVincenzo, J.~A. Smolin, Capacities of quantum erasure
  channels, Phys. Rev. Lett. 78 (1997) 3217--3220.

\bibitem{HHH+09}
R.~Horodecki, P.~Horodecki, M.~Horodecki, K.~Horodecki, Quantum entanglement,
  Rev. Mod. Phys. 81 (2009) 865--942.

\bibitem{KMN+07}
C.~King, K.~Matsumoto, M.~Nathanson, M.~B. Ruskai, Properties of conjugate
  channels with applications to additivity and multiplicativity, Markov Process
  and Related Fields 13 (2007) 391--423.

\bibitem{CG21}
S.~Chessa, V.~Giovannetti, Quantum capacity analysis of multi-level amplitude
  damping channels, Commun. Phys. 4 (2021) 22.

\bibitem{WP07}
M.~M. Wolf, D.~P\'erez-Garc\'{\i}a, Quantum capacities of channels with small
  environment, Phys. Rev. A 75 (2007) 012303.

\end{thebibliography}
\bibliographystyle{elsarticle-num}

\end{document}